\documentclass[prl,twocolumn,showpacs,superscriptaddress,amssymb,10pt]{revtex4-1}

\usepackage{graphicx}
\usepackage{dcolumn}
\usepackage{bm}
\usepackage{epsfig}
\usepackage{color}
\usepackage{longtable}
\usepackage{amsmath}
\usepackage{multirow}
\usepackage{tabularx}

\definecolor{aogreen}{rgb}{0.0, 0.5, 0.0}
\def\rmem#1#2#3{  \left\langle #1 \left\vert \left\vert  #2
                  \right\vert \right\vert #3 \right\rangle   }
\def\twobytwo#1#2#3#4{  \left( \begin{array}{cc}
                                   #1 & #2   \\[0.2cm]
                                   #3 & #4   \end{array} \right)   }

\definecolor{mymainmessagecolor}{RGB}{10,200,10}

\begin{document}

\preprint{}
\title{
Maximum Elliptical Dichroism in Atomic Two-photon Ionization\\
}

\author{J.~Hofbrucker}
\affiliation{Helmholtz-Institut Jena, Fr\"o{}belstieg 3, D-07743 Jena, Germany}
\affiliation{Theoretisch-Physikalisches Institut, Friedrich-Schiller-Universit\"at Jena, Max-Wien-Platz 1, D-07743
Jena, Germany}%

\author{A.~V.~Volotka}
\affiliation{Helmholtz-Institut Jena, Fr\"o{}belstieg 3, D-07743 Jena, Germany}%

\author{S.~Fritzsche}
\affiliation{Helmholtz-Institut Jena, Fr\"o{}belstieg 3, D-07743 Jena, Germany}%
\affiliation{Theoretisch-Physikalisches Institut, Friedrich-Schiller-Universit\"at Jena, Max-Wien-Platz 1, D-07743
Jena, Germany}

\date{\today \\[0.3cm]}

\begin{abstract}

Elliptical dichroism is known in atomic photoionization as the difference in the photoelectron angular distributions produced in nonlinear ionization of atoms by left- and right-handed elliptically polarized light. We theoretically demonstrate that the maximum dichroism $|\Delta_{\textrm{ED}}|=1$ always appears in two-photon ionization of any atom if the photon energy is tuned in so that the electron emission is dominantly determined by two intermediate resonances. We propose the two-photon ionization of atomic helium in order to demonstrate this remarkable phenomenon. The maximum elliptical dichroism could be used as a sensitive tool for analyzing the polarization state of photon beams produced by free-electron lasers. \\

\end{abstract}

\newpage
\maketitle

It is well known, that when unpolarized atoms are irradiated with ionizing \textit{circularly} polarized light, photoelectron angular distributions are identical for left and right handedness of the light. When atoms are initially oriented, however, photoelectron distributions generally depend on the handedness of the light \cite{Baum/PRL:1970, Cherepkov/JPB:1995}, and the different outcome for left- and right-polarized light, so-called \textit{circular dichroism}, has been explored for many years. Since its discovery, it has become an inevitable tool for studying biomolecules \cite{Fasman/Book:1996} and for determining the structure of chiral molecules \cite{Beaulieu/Nat:2018}. However, it has also found an application in other fields, for example in polarization effect control \cite{Mousavi/SR:2015} or optical activity control in metamaterials \cite{Khanikaev/NC:2016}.

Three decades ago, left-right asymmetries in photoelectron angular distributions in above-threshold ionization of unoriented noble gas atoms by elliptically polarized light were observed for the first time \cite{Bashkansky/PRL:1988}, and it was demonstrated that dichroism does not require a chiral target but arises also from nonlinear interactions with elliptically polarized light. Although the origin of the asymmetries remained unclear to the authors at that time, the reason for this asymmetry have been found already in an earlier work that year \cite{Kassaee/PRA:1988}, in which this asymmetry remained unnoticed. The dots were soon connected by two theoretical groups \cite{Lambropoulos/PRL:1988,Muller/PRL:1988} who provided a brief explanation of the phenomenon based on lowest-order perturbation theory. While these theories fully describe the observed elliptical dichroism, a lucid explanation was still missing until today.

In Ref. \cite{Goreslavski/PRL:2004}, for example, it was shown that asymmetries in the angular distributions can be understood from (changes in) the Coulomb potential as seen by the emitted electron. From a theoretical point of view, this means that the widely used Keldysh approximation is insufficient, and, hence, the binding potential needs to be treated explicitly \cite{Basile/PRL:1988,Jaron/OC:1999}. In the 1990s, however, such studies of the elliptical dichroism were restricted by the low energies of available lasers. These lasers only allowed two schemes to be realized, either production of slow photoelectrons due to absorption of the minimal number of photons necessary for ionization to occur, or by making use of above-threshold ionization by absorbing additional photons in the focus of strong laser fields \cite{Paulus/PRL:2000, Paulus/PRL:1998, Manakov/JPB:1999}. The first option was performed experimentally, e.g., for the two-photon ionization of the rubidium $5s~^2S_{1/2}$ electron \cite{Wang/PRL:2000, Wang/PRA:2000}, with emphasis on extracting relative phases and transition amplitudes from the photoelectron angular distributions \cite{Dulieau/JPB:1995}. This experiment gave rise to an unexpected cross section ratio of the two fine structure channels (partial \textit{d} wave), and suggested much stronger spin-orbit effects than predicted theoretically \cite{Colgan/PRL:2001}. Further studies of elliptical dichroism were performed for near-threshold energies and confirmed strong asymmetries for several atoms \cite{Borca/PRL:2001, Hofbrucker/PRA:2016, Hofbrucker/PRA:2017}. However, it is typically difficult to estimate the magnitude of the dichroism parameter for these near-threshold energies without performing detailed calculations.

The rise of free-electron lasers (FELs) during the last two decades removed the restrictions of optical lasers and opened up new possibilities for studying nonlinear processes. In particular, polarization control of intense high-energy FEL pulses \cite{Allaria/NatPhot:2013,Lutman/NatPhot:2016} enables one today to test the chiral or dichroic properties of matter with unprecedented accuracy \cite{Kazansky/PRL:2011,Meyer/PRL:2008} and in extreme ultraviolet or x-ray energy domains. For example, Ilchen \textit{et al.} \cite{Ilchen/PRL:2017} have recently studied circular dichroism of oriented He$^{+}$ ions and found the atomic orientation imprinted on both, the differential and total cross sections. On the other, circular dichroism can also be used as a tool for analyzing the polarization state of FELs \cite{Mazza/NatCom:2014}. Further investigations with highly energetic circularly or elliptically polarized photons from FELs may, therefore, help improve our understanding of elliptical dichroism and chirality \cite{Seipt/PRA:2016,Ilchen/PRA:2017} and will open new applications in atomic and molecular physics.

In this Letter, we theoretically demonstrate that for two-photon ionization of atomic targets, it is always possible to detect maximum elliptical dichroism ($|\Delta_{\textrm{ED}}|=1$) for any atom and for properly tuned photon energy. In particular, a maximum elliptical dichroism is achieved if the virtual state that the atoms take after absorption of the first photon, is sandwiched between two resonances. We also show that this maximum left-right asymmetry in the photoelectron angular distribution from atomic two-photon ionization by elliptically polarized light can be readily understood geometrically from the properties of spherical harmonics and photon polarization. The strong dichroic response of atoms to elliptically polarized light could be utilized for detailed analysis in the polarization control of FEL beams. 

\begin{figure}[t]
\includegraphics[scale=0.32 ]{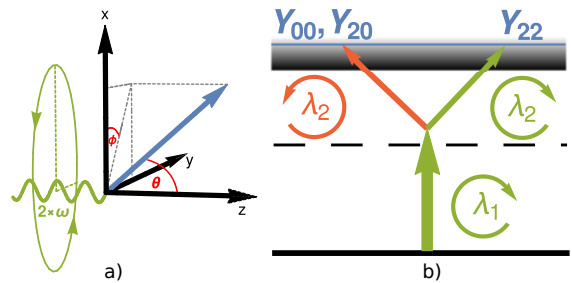}
\caption{(a) The incident photons propagate along the quantization $z$ axis, the polarization major semiaxis is along the $x$ axis, and the electron emission direction is given by two angles $\theta$ and $\phi$.
(b) Angular momentum scheme for the two-photon ionization of an \textit{s}-state electron. The selection rules provide a simple relationship between the photon helicities $\lambda$ and the final symmetries of the photoelectron partial waves. }\label{Fig.Channels}
\end{figure}

Elliptical dichroism arises already within nonrelativistic electric dipole approximation \cite{Manakov/JPB:1999,Hofbrucker/PRA:2017} and without any need to resort to the electron spin. Let us consider an \textit{s} electron in a bound (atomic) state and elliptically polarized photons described by a wave vector $\textbf{k}$, and with polarization state given by a $2\times2$ density matrix \cite{Blum/Book:1981}
\begin{eqnarray}
\rho_{\gamma}^{\lambda \lambda'}=\frac{1}{2}\twobytwo{1+P_c}{P_l}{P_l}{1-P_c},
\end{eqnarray}
where $P_l$ and $P_c$ denote the linear and circular Stokes parameters. After the interaction of the \textit{s} electron with both photons, the electron either undergoes an \mbox{$s \rightarrow p \rightarrow s$} or \mbox{$s \rightarrow p \rightarrow d$} transition, and it is released eventually with kinetic energy $\varepsilon$ into some direction $\theta$ and $\phi$ (see Fig. \ref{Fig.Channels}). The photoelectron is assumed to be in a pure state, and its wave function can be simply written as a sum of two partial waves
\begin{equation}\label{Eq.wavefunction}
\psi_{\varepsilon}(\textbf{r})=a_0\psi_{\varepsilon00}(\textbf{r})+\sum_{m_d} b_{m_d}\psi_{\varepsilon2m_d}(\textbf{r})
\end{equation}
using the usual notation \mbox{$\psi_{\varepsilon l m}(\textbf{r})=\psi_{\varepsilon l}(r)Y_{lm}(\theta,\phi)$}, and where $m_d = 0, \pm2$, since the quantization axis is chosen along the photon propagation direction. The \textit{s} and \textit{d} partial-wave amplitudes $a_0$ and $b_{m_d}$, respectively, contain all information about the dynamics of the ionization process, 
\begin{equation}\label{Eq.abrho}
a_0, b_{m_d} \propto \rho_\gamma^{\lambda_1\lambda_1'} \rho_\gamma^{\lambda_2\lambda_2'},
\end{equation}
and their exact expressions can be obtained similarly from the photon density matrix as in Ref. \cite{Dulieau/JPB:1995}. 
The propagation direction of the photoelectron is completely characterized by its probability density
\begin{eqnarray}\label{Eqn.ProbabilityDensity}
|\psi_{\varepsilon}|^2	&=&	
 |a_0\psi_{\varepsilon00}|^2
+|\sum_{m_d} b_{m_d}\psi_{\varepsilon2m_d}(\textbf{r})|^2\nonumber \\ 
&+&\sum_{m_d}a_0^*b_{m_d}\psi^*_{\varepsilon00}(\textbf{r})\psi_{\varepsilon2m_d}(\textbf{r})\\
&+&\sum_{m_d}a_0b_{m_d}^*\psi_{\varepsilon00}(\textbf{r})\psi^*_{\varepsilon2m_d}(\textbf{r}).\nonumber
\end{eqnarray}
We shall analyze this expression in order to explore the left-right asymmetry in photoelectron angular distributions in further detail. Mathematically speaking, this means that we are looking for an antisymmetric contribution in Eq. (\ref{Eqn.ProbabilityDensity}), which changes its sign under the coordinate transformation $y\rightarrow -y$, or, equivalently $\phi \rightarrow -\phi$. Since the angular dependence of the wave function is described by the spherical harmonics, we easily see that only the imaginary part of spherical harmonics $\textrm{Im}[Y_{2\pm2}(\theta, \phi)]$ changes its sign with the $\phi \rightarrow -\phi$ transformation. Since this term is complex,  we conclude that the antisymmetric contribution must be contained only in the interference terms. 
In the dipole approximation, the well-known selection rules provide a simple relationship between the helicities $\lambda_1$ and $\lambda_2$ ($\lambda_1'$ and $\lambda_2'$) of the two photons for obtaining a particular projection of the angular momentum $m$ of each partial wave. For example, in order to obtain $m=\pm 2$, both photons must have the same helicity $\lambda_1=\lambda_2$, while we must have $\lambda_1=-\lambda_2$ for $m=0$ (see Fig. \ref{Fig.Channels} for graphical representation). We, therefore, expect, the $\phi$-dependent part of the interference to be proportional to $\propto \sum_{\lambda_1 \lambda_1'}Y_{0 \lambda_1-\lambda_1} Y^*_{2 \lambda_1'+\lambda_1'} \rho_\gamma^{\lambda_1 \lambda_1'}\rho_\gamma^{\lambda_1 -\lambda_1'}\propto iP_l P_c \sin (2 \phi)$. This is exactly the antisymmetric term responsible for the dichroism which changes sign under the $\phi \rightarrow -\phi$ transformation or equivalently changes sign upon a change of photon handedness. Moreover, we see that for pure linear or circular polarization ($P_c=0$ or $P_l=0$) this term vanishes. A similar analysis shows, that there is no elliptical dichroism in photoelectron angular distribution of one-photon ionization of a spherically symmetric target and, hence, that elliptical dichroism is purely a nonlinear phenomenon. 

\begin{figure}[t]
\includegraphics[scale=0.35]{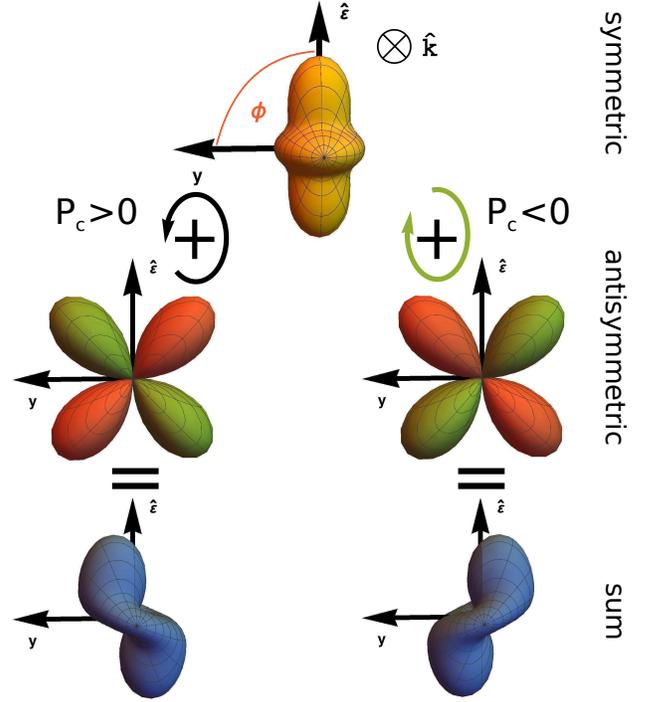}
\caption{The origin of elliptical dichroism in the photoelectron angular distribution of two-photon ionization of an \textit{s} electron. We can separate the distribution into a symmetric core contribution (top) and an antisymmetric dichroic contribution (middle row). The dichroic part is given by a $\sin (2\phi)$ and, therefore, has positive (green) and negative (red) intervals. The sum of the core and dichroic contributions gives us the final photoelectron distribution (bottom row). }\label{Fig.3dall}
\end{figure}

The geometrical analysis given above predicts the presence of the dichroism. However, the photoelectron angle-differential cross section can be certainly derived rigorously. This has been done previously for two-photon ionization of \textit{s}-state electrons for arbitrary polarization \cite{Manakov/JPB:1999, Hofbrucker/PRA:2017} the cross section is given by
\begin{eqnarray}\label{Eq.CrossSection}
\frac{d\sigma}{d\Omega}&=&
\frac{9 \pi^2 \alpha^2}{2 \omega^2} \Big \{
|U_s|^2 \mathcal{P}+|U_d|^2\Big[ \mathcal{P}-3\mathrm{sin}^2\theta \Big(\mathcal{P}+\nonumber\\
&+&2P_l\mathrm{cos}(2\phi)\Big)+\frac{9}{2}\mathrm{sin}^4\theta\Big(1+P_l\mathrm{cos}(2\phi)\Big)^2 \Big]+\nonumber \\
&+&2\mathrm{Re}\Big[U_s U_d^*e^{i(\delta_s-\delta_d)}[ \mathcal{P}-\frac{3}{2}\mathrm{sin}^2\theta \big(\mathcal{P} + \nonumber \\
&+&2P_l\mathrm{cos}(2\phi)+2iP_l P_c \mathrm{sin}(2\phi)\big)\big]\Big]\Big\},
\end{eqnarray}
with \mbox{$\mathcal{P}=1+P_l^2-P_c^2$}, partial-wave phases $\delta_{s,d}$, and radial transition amplitudes $U_{s,d}$. We can decompose the differential cross section into symmetric and antisymmetric contributions (see Fig. \ref{Fig.3dall} for visualization). The "core" symmetric part (orange) contains the squared terms of Eq.~(\ref{Eqn.ProbabilityDensity}) and the symmetric part of the interference term. The last term of Eq. (\ref{Eq.CrossSection}) represents the "dichroic" asymmetric part (red and green), and it is the only term depending on the photon handedness. The signs of the circular Stokes parameter $P_c$ and the phase difference $\delta'\equiv\delta_{s}-\delta_{d}$ determine the intervals for which the dichroic term contributes constructively (green) or destructively (red). The sum of the core and the dichroic contributions gives the final photoelectron angular distribution (blue). The relative contributions of the core and dichroic parts, and hence the magnitude of the left-right asymmetry, are consequently determined by the ratio of the partial waves $u\equiv U_d/U_s$. The asymmetry in photoelectron angular distribution can then be quantified by introducing a dichroism parameter $\Delta_{\textrm{ED}}$ defined as
\begin{eqnarray}\label{Eq.Dichroism}
\Delta_{\mathrm{ED}}(\theta, \phi)=\frac{d\sigma_{+}/d\Omega-d\sigma_{-}/d\Omega}{d\sigma_{+}/d\Omega+d\sigma_{-}/d\Omega},
\end{eqnarray}
with the index $+/-$ corresponding to the sign of $P_c$. The dichroism parameter takes values in the interval $\Delta_{\textrm{ED}} \in [-1,1]$, where $|\Delta_{\textrm{ED}}|=1$ describes the maximum possible effect. Our aim is to find conditions under which this maximum can be detected. Although Eq. (\ref{Eq.CrossSection}) describes also the above-threshold ionization \cite{Manakov/JPB:1999}, in this Letter, we will solely discuss the case, where the photon energy is lower than the one-photon ionization threshold.

\begin{figure*}[t]
\includegraphics[scale=0.6]{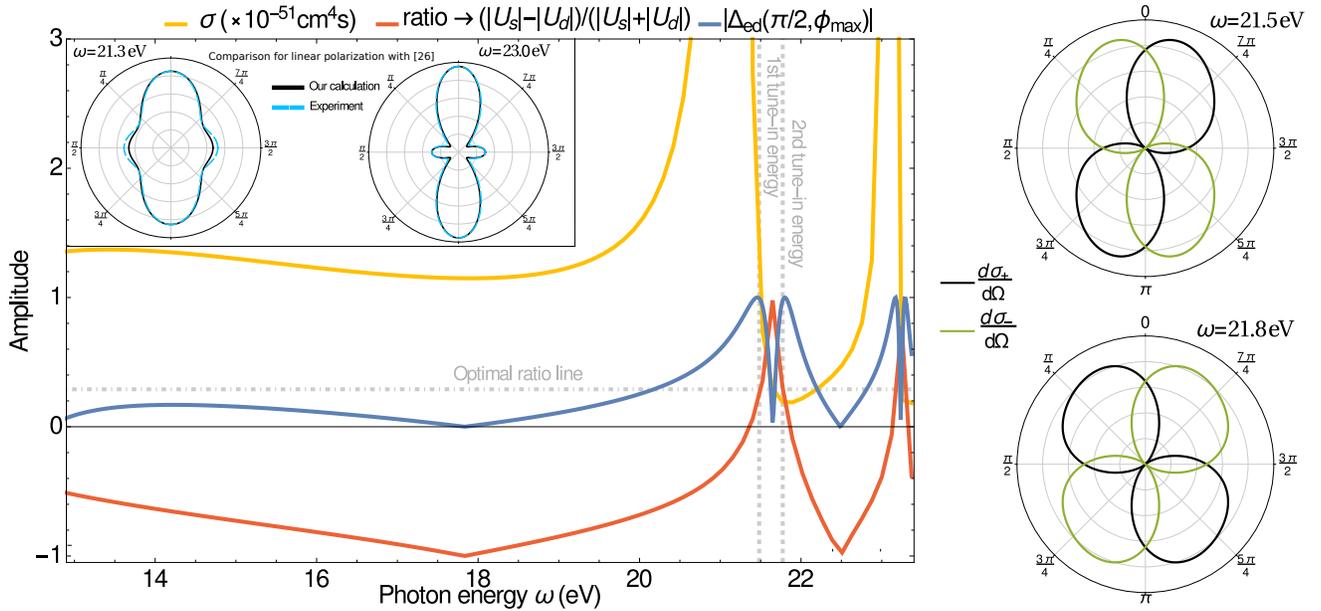}
\caption{Total  cross section (yellow), partial-wave ratio (red), and elliptical dichroism parameter (blue) as functions of single photon energy. The horizontal gray dot-dashed line signifies the optimal ratio for obtaining $|\Delta_{\textrm{ED}}|=1$. This line intersects the partial-wave ratio at two tune-in energies (marked by two gray dotted vertical lines). Photoelectron angular distributions corresponding to these two intersections are shown on the right side of the figure. Inset in the top left corner shows the angular distribution comparisons with an experiment \cite{Ma/JPB:2013} for two-photon ionization of He by linearly polarized light.}\label{Fig.He_master}
\end{figure*}

To find the maximum dichroism, we concentrate on distributions in the polarization plane (perpendicular to the photon propagation direction, i.e., $\theta=\pi/2$). Moreover, since the FEL pulses possess a high degree of polarization, we can consider a fully polarized beam which is half linearly and half circularly polarized, $P_l^2=1/2$ and $P_c^2=1/2$. We can insert Eq. (\ref{Eq.CrossSection}) into Eq. (\ref{Eq.Dichroism}) to obtain the expression for elliptical dichroism parameter. By analyzing the second derivatives of $\Delta_{\textrm{ED}}$, we find that the dichroism parameter reaches its extrema ($+1$ or $-1$) at particular azimuthal angles $\phi_{\textrm{max}}$, and corresponding values of the amplitude ratio \mbox{$u_{\textrm{max}}=2[(3\cos (2\phi_{\textrm{max}})+\sqrt{2})^2+8]^{-1/2}$}. The angle $\phi_{\textrm{max}}$ can be obtained from a fit to numerical solutions and is given by \mbox{$\phi_{\textrm{max}}(\delta')=0.95 \delta' - 0.33 \delta'^2 +  0.06 \delta'^3$} for $0<\delta'<\pi$, and $\phi_{\textrm{max}}(2\pi-\delta')$ for $\pi<\delta'<2\pi$. Note that at this angle either $d\sigma_{+}/d\Omega=0$ or $d\sigma_{-}/d\Omega=0$. The elliptical dichroism at this angle can, therefore, be used for sensitive extraction of the phase information and amplitude ratios. The important conclusion is that for any nonzero phase difference, there exists an amplitude ratio for which the elliptical dichroism reaches its maximum. Of course, we cannot dictate nature what the amplitude ratio should be. We can, nevertheless, search for photon energies $\omega$, for which the transition amplitudes fulfill the above condition (assuming that $\delta'$ varies much slower than $u$). By tuning the photon energy such that the virtual intermediate state lies between two resonances, the transition amplitudes of the partial waves are dominantly determined by   
\begin{eqnarray}\label{Eqn.IntermediateTA}
U_{l} &\propto& \frac{\rmem{\alpha_f J_f \varepsilon l}{r}{\alpha_{1} J_{1}}\rmem{\alpha_{1} J_{1}}{r}{\alpha_i J_i}}{E_i+\omega-E_{1}}\\\nonumber 
&+&\frac{\rmem{\alpha_f J_f \varepsilon l}{r}{\alpha_{2} J_{2}}\rmem{\alpha_{2} J_{2}}{r}{\alpha_i J_i}}{E_i+\omega-E_{2}},
\end{eqnarray}
with quantum numbers $\alpha$, angular momenta $J$, the ground, and intermediate resonance state energies $E_i$, $E_{1}$, and $E_2$, respectively. Note that so far no particular electron shell or atom was specified. For certain photon energy $\omega$, these two contributions cancel out, and the transition amplitude becomes zero (similar for the tune-out wavelength \cite{Arora/PRA:2011}). Since the amplitudes are continuous functions of energy and drop to zero for different photon energies, there are always two "tune-in" photon energies $\omega_{\textrm{max}} \in ( E_1-E_i, E_2-E_i )$, which guarantee the fulfillment of the optimal ratio $u_{\textrm{max}}$. 

Before we demonstrate our findings with an example, we would like to draw a brief conclusion. Maximum elliptical dichroism in the photoelectron angular distribution can be obtained by tuning in the photon energy $\omega_{\textrm{max}}$ so that after absorption of one photon, an $s$ electron of an atom is promoted to an intermediate virtual state between $np$ and $(n+1) p$ resonances, with the exception of high $n$ states, for which the corresponding width is comparable to the energy separation of the resonances. To demonstrate the generality of our findings, we show  in the Supplementary Material \citep{SP}, that maximum elliptical dichroism is present in two-photon ionization of all electron shells of a Ca atom.


We wish to demonstrate detection of maximum elliptical dichroism on an experimentally plausible example of two-photon ionization of ground state helium, which has already been used as a target for various atomic studies in the past decade \cite{Meyer/PRL:2008,Kazansky/PRL:2011,Ilchen/PRL:2017}, and where two-photon ionization is the dominant process. Moreover, an experiment conducted at the SPring-8 facility used the photoelectron angular distributions of two-photon ionization of helium by linearly polarized light to extract transition amplitudes and phase-shift differences \cite{Ma/JPB:2013}. This work allowed us to compare our theoretical treatment with an experiment. We base our theory on second-order perturbation theory and independent particle approximation\cite{Hofbrucker/PRA:2016, Hofbrucker/PRA:2017, Surzhykov/JPB:2015}. We reproduce the experimental results with a perfect agreement (see the inset of Fig.~\ref{Fig.He_master}), which confirms that our theory is suitable for describing the reported effect.

To obtain maximum dichroism in two-photon ionization of He, we need to tune in a photon energy which promotes one of the $1s$ electrons into continuum through a virtual intermediate state sandwiched between two resonances: $1s^2\rightarrow1s2p$ and $1s^2\rightarrow1s3p$. The resonances are clearly visible in the plot of the total ionization cross section in Fig. \ref{Fig.He_master} (yellow). The red plot represents the normalized ratio of the partial waves $(|U_s|-|U_d|)/(|U_s|+|U_d|)\in [-1, 1]$. According to the propensity rules \cite{Fano/PRA:1985}, upon absorption of a photon, transitions corresponding to $l\rightarrow l+1$ should be favored, and the normalized partial-wave ratio should be negative. Figure \ref{Fig.He_master} shows that this is generally true; however, around the resonances, the ratio strongly deviates from these results. The amplitude ratio becomes positive at a point, where the dominant contributions from Eq. (\ref{Eqn.IntermediateTA}) cancel each other out, and it reaches unity when $U_d=0$. Between these two significant points, the amplitude ratio passes the optimal ratio (horizontal gray line), where maximum elliptical dichroism can be observed. After a phase jump induced by transition of $U_d$ through zero, the amplitude ratio decreases and passes the optimal ratio again. The fulfillment of the optimal ratio are also clearly visible in the blue plot of the $\Delta_{\textrm{ED}}(\pi/2, \phi_{\textrm{max}}$) as a function of the photon energy, with the two peaks between the $2p-3p$ resonances (and another two between the $3p-4p$ resonances), each representing the maximum possible dichroism. The corresponding photoelectron angular distributions for these maxima are provided on the right-hand side of Fig. \ref{Fig.He_master} in the polarization plane ($\theta=\pi/2$) for both left-(green) and right-handed (black) elliptically polarized light. In a real experiment, these distributions could be influenced by the bandwidth of the incoming laser beam. However, for He, the width would be lower than $<0.1$ eV. Our calculations show that such an energy deviation would not significantly influence the photoelectron distributions.

Although two-photon ionization of helium was already performed for photon energies where elliptical dichroism is strong, it was carried out only with linearly polarized light \cite{Ma/JPB:2013}. Nevertheless, the interest in polarization control at FEL facilities is growing. Currently, there are already two FEL facilities able to produce elliptically polarized beams: FERMI using their Apple~II undulator \cite{Allaria/NatPhot:2013} and LCLS using the Delta undulator \cite{Lutman/NatPhot:2016}. These two FELs can produce elliptically polarized beams only in high-energy ranges (24-120~eV at FERMI and 500-1200~eV at LCLS); however, there are numerous upgrade plans at other facilities to control the light polarization. 

In conclusion, we showed that maximum elliptical dichroism ($|\Delta_{\textrm{ED}}|=1$) can always be detected in two-photon ionization of \textit{s} electrons for a system specific photon energy. The results were demonstrated for the case of two-photon ionization of helium but are generally applicable to the ionization of \textit{s}-state electrons of arbitrary atoms. Two-photon ionization, therefore, provides a unique opportunity to access electronic potential properties and extract dynamic information about the atomic or molecular ionization process. The reported phenomenon could be used also as a fine tool to analyze the polarization state (or ellipticity) of FEL pulses, similar to Ref. \cite{Mazza/NatCom:2014}, and to accurately extract the amplitude ratio and phase differences from the electron angular distributions.

\begin{acknowledgments}
We greatly appreciate the fruitful discussions with M.~Ilchen, and M.~Meyer, which enlightened us on the current experimental possibilities. This work has been supported by the BMBF (Grant No. 05P15SJFAA).
\end{acknowledgments}

%
%
%


\begin{thebibliography}{28}
\bibitem{Baum/PRL:1970} G. Baum, M. S. Lubell, and W. Raith, Phys. Rev. Lett. \textbf{25}, 267 (1970).
\bibitem{Cherepkov/JPB:1995} N. A. Cherepkov, V. V. Kuznetsov, and V. A. Verbitskii, J. Phys. B \textbf{28}, 1221 (1995).
\bibitem{Fasman/Book:1996} G. D. Fasman, \textit{Circular Dichroism and the Conformational Analysis of Biomolecules} (Springer, New York, 1996).
\bibitem{Beaulieu/Nat:2018} S. Beaulieu, A. Comby, D. Descamps, B. Fabre, G. A. Garcia, R. Géneaux, A. G. Harvey, F. Légaré, Z. Mašín, L. Nahon, A. F. Ordonez, S. Petit, B. Pons, Y. Mairesse, O. Smirnova, and V. Blanchet, Nat. Phys. \textbf{14}, 484 (2018).
\bibitem{Mousavi/SR:2015} S.A. Mousavi, E. Plum, J. Shi, and N. I. Zheludev, Sci. Rep. \textbf{5}, 8977 (2015). 
\bibitem{Khanikaev/NC:2016} A. B. Khanikaev, N. Arju, Z. Fan, D. Purtseladze, F. Lu, J. Lee, P. Sarriugarte, M. Schnell, R. Hillenbrand, M. A. Belkin, and G. Shvets, Nat. Commun. \textbf{7}, 12045 (2016).
\bibitem{Bashkansky/PRL:1988} M. Bashkansky, P. H. Bucksbaum, and D. W. Schumacher, Phys. Rev. Lett. \textbf{60}, 2458 (1988).
\bibitem{Kassaee/PRA:1988} A. Kassaee, M. L. Rustgi, and S. A. T. Long, Phys. Rev. A \textbf{37}, 999 (1988).
\bibitem{Lambropoulos/PRL:1988} P. Lambropoulos, and X. Tang, Phys. Rev. Lett. \textbf{61}, 2506 (1988).
\bibitem{Muller/PRL:1988} H. G. Muller, G. Petite, and P. Agostini, Phys. Rev. Lett. \textbf{61}, 2507 (1988).
\bibitem{Goreslavski/PRL:2004} S. P. Goreslavski, G. G. Paulus, S.V. Popruzhenko, and N. I. Shvetsov-Shilovski, Phys. Rev. Lett. \textbf{93}, 233002 (2004).
\bibitem{Basile/PRL:1988} S. Basile, F. Trombetta, and G. Ferrante, Phys. Rev. Lett. \textbf{61}, 2435 (1988).
\bibitem{Jaron/OC:1999} A. Jaro\'n \textit{et al.}, Opt. Commun. \textbf{163}, 115 (1999).
\bibitem{Paulus/PRL:1998} G. G. Paulus , F. Zacher, H. Walther, A. Lohr, W. Becker, and M. Kleber, Phys. Rev. Lett. \textbf{80}, 484 (1998).
\bibitem{Paulus/PRL:2000} G. G. Paulus, F. Grasbon, A. Dreischuh, H. Walther, R. Kopold, and W. Becker, Phys. Rev. Lett. \textbf{84}, 3791 (2000).
\bibitem{Manakov/JPB:1999} N. L. Manakov, A. Maquet, S. I. Marmo, V. Veniard, and G. Ferrante, J. Phys. B \textbf{32}, 3747 (1999).
\bibitem{Wang/PRL:2000} Z. M. Wang and D. S. Elliott, Phys. Rev. Lett. \textbf{84}, 3795 (2000).
\bibitem{Wang/PRA:2000} Z. M. Wang and D. S. Elliott, Phys. Rev. A \textbf{62}, 053404 (2000).
\bibitem{Dulieau/JPB:1995} F. Dulieu, C. Blondel, and C. Delsart, J. Phys. B \textbf{28}, 3845 (1995).
\bibitem{Colgan/PRL:2001} J. Colgan, and M. S. Pindzola, Phys. Rev. Lett. \textbf{86}, 1998 (2001).
\bibitem{Borca/PRL:2001} B. Borca, M. V. Frolov, N. L. Manakov, and A. F. Starace, Phys. Rev. Lett. \textbf{87}, 133001 (2001).
\bibitem{Hofbrucker/PRA:2016} J. Hofbrucker, A. V. Volotka, and S. Fritzsche, Phys. Rev. A \textbf{94}, 063412 (2016).
\bibitem{Hofbrucker/PRA:2017} J. Hofbrucker, A. V. Volotka, and S. Fritzsche, Phys. Rev. A \textbf{96}, 013409 (2017).
\bibitem{Allaria/NatPhot:2013} E. Allaria \textit{et al.}, Nat. Photonics \textbf{7}, 913 (2013).
\bibitem{Lutman/NatPhot:2016} A. A. Lutman \textit{et al.}, Nat. Photonics \textbf{10}, 468 (2016).
\bibitem{Meyer/PRL:2008} M. Meyer \textit{et al.} Phys. Rev. Lett. \textbf{101}, 193002 (2008).
\bibitem{Kazansky/PRL:2011} A. K. Kazansky, A. V. Grigorieva, and N. M. Kabachnik, Phys. Rev. Lett. \textbf{107}, 253002 (2011).
\bibitem{Ilchen/PRL:2017} M. Ilchen \textit{et al.}, Phys. Rev. Lett. \textbf{118}, 013002 (2017).
\bibitem{Mazza/NatCom:2014} T. Mazza \textit{et al.}, Nat. Commun. \textbf{5}, 3648 (2014).
\bibitem{Seipt/PRA:2016} D. Seipt, R. A. M\"uller, A. Surzhykov, and S. Fritzsche, Phys. Rev. A \textbf{94}, 053420 (2016).
\bibitem{Ilchen/PRA:2017} M. Ilchen \textit{et al.}, Phys. Rev. A \textbf{95}, 053423 (2017).
\bibitem{Blum/Book:1981} K. Blum, \textit{Density Matrix Theory and Applications} (Plenum, New York, 2000).
\bibitem{Arora/PRA:2011} B. Arora, M. S. Safronova, and C. W. Clark, Phys. Rev. A \textbf{84}, 043401 (2011).
\bibitem{SP} See Supplemental Material of Phys. Rev. Lett. \textbf{121}, 053401 (2018), for further details.
\bibitem{Ma/JPB:2013} R. Ma \textit{et al.}, J. Phys. B \textbf{46}, 164018 (2013).
\bibitem{Surzhykov/JPB:2015} A. Surzhykov, V. A. Yerokhin, Th. St\"ohlker, and S. Fritzsche, J. Phys. B \textbf{48}, 144015 (2015).
\bibitem{Fano/PRA:1985} U. Fano, Phys. Rev. A \textbf{32}, 617 (1985).

\end{thebibliography}
\end{document}